\documentclass
[aps,twocolumn,showpacs,groupedaddress,amsmath,floatfix,nofootinbib]{revtex4}

\advance\textheight by 0.25in
\advance\voffset by -0.125in

\usepackage{graphicx}
\usepackage{epsfig}
\usepackage{amsfonts}
\usepackage{amsmath,amssymb}
\usepackage{color}

\newcommand{\be}{\begin{eqnarray}}
\newcommand{\ee}{\end{eqnarray}}

 \newcommand{\GeV}{\hbox{GeV}}
 \newcommand{\tw}{\textwidth}
 \newcommand{\ave}[1]{\langle #1 \rangle}
\begin{document}


\title{Charge Transfer Fluctuations as a QGP Signal}

\author{Sangyong Jeon}
\email[] {jeon@physics.mcgill.ca}
\affiliation{Physics Department, McGill University,
Montr{\'e}al Qu\'ebec H3A 2T8, Canada}
\affiliation{RIKEN-BNL Research Center, Upton NY 11973, USA}
\author{Lijun Shi}
\affiliation{Physics Department, McGill University,
Montr{\'e}al Qu\'ebec H3A 2T8, Canada}
\author{Marcus Bleicher}
\affiliation{Institut f\"{u}r Theoretische Physik,
 Johann Wolfgang Goethe -- Universit\"{a}t,
 Robert Mayer Str. 8-10,
 60054 Frankfurt am Main, Germany}
\date{\today}

\begin{abstract}
In this study, we analyze the recently proposed
charge transfer fluctuations within a finite pseudo-rapidity space.
As the charge transfer fluctuation is a measure of the local charge
correlation length,
it is capable of detecting inhomogeneity in the hot and dense
matter created by heavy ion collisions.
We predict that going from peripheral to central collisions,
the charge transfer fluctuations at midrapidity should decrease
substantially while
the charge transfer fluctuations at the
edges of the observation window should decrease by a small amount.
These are consequences of having a strongly inhomogeneous matter
where the QGP component is concentrated around midrapidity.
We also show how to constrain the values of the charge
correlations lengths in both the hadronic phase and the QGP phase using
the charge transfer fluctuations.
Current manuscript is based on the preprints
hep-ph/0503085 (to appear in Physical Review C) and nucl-th/0506025.
\end{abstract}

\maketitle

\section{Introduction \label{sect::intro}}

%
%

In the last few decades, one of the major goals of the high energy nuclear
physics community has been creating and studying Quark Gluon Plasma (QGP). 
Many observables have been suggested and studied as QGP signals. Yet, there
still is no consensus on the so-called `smoking-gun' signals even though may
physicists, including myself, have a strong conviction that the strong
jet-quenching and the strong elliptic flow could only come from a QGP.
Possible reasons for such a situation may be 
that some of the proposed QGP signals didn't work that well and/or
it turned out that they still had a possible hadronic explanation.
Hence, it is crucial that we characterize the created QGP in as many ways 
as we can {\em and} figure out why some of the promising signals didn't work
that well.

In this article, we propose that the charge transfer fluctuation is a
good candidate to accomplish just that -- characterize a QGP in a new way
and at the same time provide an answer to why some signals, especially the
net charge fluctuation, didn't work as well as expected.
We will demonstrate below that by measuring
charge transfer fluctuations as a function of (pseudo-)rapidity, one can
not only detect the presence of a QGP but also 
can estimate the {\em size} of the created QGP.
We will also show that current data on net charge fluctuations 
(which is not the same as charge transfer fluctuations)
is consistent with having only about 20\,\% of all final
state particles remembering its QGP origin 
(about 50\,\% within $-1 < \eta < 1$).
This amount is small enough that any signal that needs averaging
over a large rapidity interval will be masked by the hadronic part. 

\begin{figure}[t]
\centerline{\includegraphics[angle=90,width=0.4\tw]{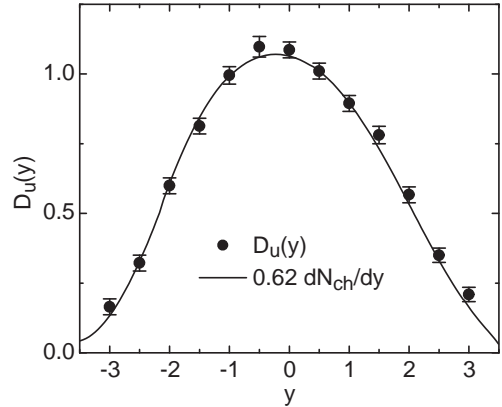}}
\caption{
Comparison between $D_u(y)$ and $dN_{\rm ch}/dy$ in pp collisions.
Kafka et.al. PRL 34, 687, 1975, $p_{\rm max} = 200\,\GeV$}
\label{fig:kafka}
\end{figure}

The charge transfer fluctuation is defined by
\be
D_u(y) = \ave{u(y)^2} - \ave{u(y)}^2
\ee
where the charge transfer $u(y)$ at rapidity $y$ is defined by
\be
u(y) = [Q_F(y) - Q_B(y)]/2
\label{eq:uy}
\ee
where
\be
\left\{
\begin{array}{l}
Q_F(y) = \hbox{Net charge in the forward region of $y$}
\\
Q_B(y) = \hbox{Net charge in the backward region of $y$}
\end{array}
\right.
\ee
and our observable is
\be
\kappa(y) 
\equiv {D_u(y)}/(dN_{\rm ch}/dy)
\ee
This observable
is first introduced in Refs.\cite{Quigg:1973wy,Chao:1973jk}
and shown to be constant in elementary collisions as shown in
Fig.~\ref{fig:kafka}.

We will argue
below that $\kappa(y)$ is actually a measure of the {\em local}
unlike-sign charge correlation length.
Therefore if a QGP is created only in a small region around midrapidity,
$\kappa(y)$ at midrapidity
should vary the most as the centrality changes
while $\kappa(y)$ at larger $y$ should stay at an almost
constant value.
Especially, since we expect the charge correlation length in a QGP to be
much smaller than the charge correlation length in a hadronic matter
\cite{Jeon:2000wg,Asakawa:2000wh,Bass:2000az}, 
we should see $\kappa(0)$ {\em dropping} faster than $\kappa(y)$
at any other $y$ as the collisions become more central. 
On the other hand, if a QGP is not created at any centrality, 
$\kappa(y)$ should be a constant function just as it is in elementary
collisions\cite{Kafka:1975cz}.

\begin{figure}[t]
\centerline{\includegraphics[width=0.4\tw]{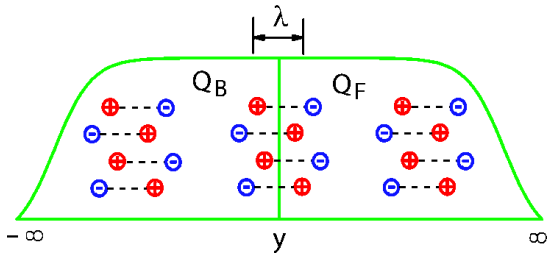}}
\caption{A schematic view of charge transfer fluctuations.
In this view, all charged particles originate from neutral clusters which
emits one positively charged particle and one negatively charged particle. 
The typical rapidity distance between the daughters is $\lambda$.
}
\label{fig:concept}
\end{figure}

\section{Charge Transfer Fluctuations}

To have a simple physical picture of the charge transfer fluctuation,
suppose that all final state charged 
hadrons originate from neutral clusters
as shown in Fig.~\ref{fig:concept}.  
Let $\lambda$ be the typical rapidity distance between the decay particles.
If a cluster decays too far from $y$, its contribution to $u(y)$ vanishes
since both of the decay products are 
either in front of or in the back of $y$.
The only time $u(y)$ is non-zero is when a neutral cluster decays
within $\lambda$ of $y$. Whenever one such neutral cluster decays, $u(y)$
undergoes a random walk with a unit step size.
Hence if there are $n$ clusters near $y$, 
\be
{D_u(y)} = n \approx \lambda{dN_{\rm cl}\over dy}
\ee
where $dN_{\rm cl}/dy$ is the rapidity density of the clusters at $y$.
Since the charged particle $dN_{\rm ch}/dy$ 
should be proportional to $dN_{\rm cl}/dy$, we then have
\be
\kappa(y) \equiv 
{{D_u(y)}\over {dN_{\rm ch}/dy}}
\propto \lambda
\ee
Having neutral particles included in cluster decays does not affect the
qualitative part of this argument.

The argument given above is essentially local.
Hence the quantity $\kappa(y)$
depends only on the properties of the local clusters. 
It is somewhat surprising that in elementary particle collisions,
$\kappa(y)$ is
actually constant as shown in Fig.~\ref{fig:kafka}.
It is also constant in non-QGP
models of nucleus-nucleus collisions such as HIJING 
(see Ref.\cite{Shi:2005rc}).
On the other hand, if a QGP is created at midrapidity, the plot of
$\kappa(y)$ should show a `dip' at $y=0$ and the depth and the
width of the dip should be an indication of the size of the QGP, or at least
the portion of hadrons that remember their QGP origin.

\section{Simple Neutral Cluster Models}

To see the effect of a QGP drop near midrapidity, we need a model.
Here we use a neutral cluster model
similar to the old $\rho,\omega$ model  \cite{Chao:1973jk} 
and Bialas et.al.'s model \cite{Bialas:1974bk}.

\begin{figure}[t]
\centerline{\includegraphics[width=0.4\tw]{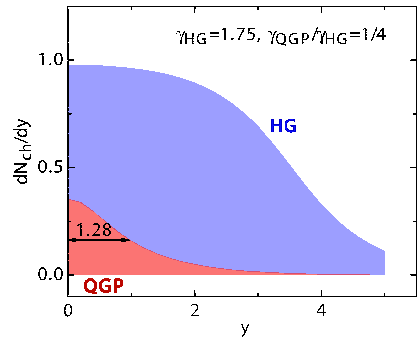}}
\caption{Schematic view of typical final state particle distributions.}
\label{fig:twocomp-profile}
\end{figure}

The procedure is as follows.
To create a simulated event with $2M_0$ charged hadrons, 
sample 
\be
{\rho_{HG}(y_+, y_-) = R_{HG}(y_+ - y_-|Y)F_{HG}(Y)}
\ee
$(1-p)M_0$ times and sample
\be
{\rho_{QGP}(y_+, y_-) = R_{QGP}(y_+ - y_-|Y)F_{QGP}(Y)}
\ee
$pM_0$ times.  From these events, obtain $\kappa(y)$.
Here $p$ is the fraction of the hadrons that remember their
QGP origin.
The function
$R(y|Y) = {1\over 2\gamma} e^{-|y|/\gamma}$ represents the correlation
between the daughters with $\gamma = 2\kappa$ and $F(Y)$ represents the
rapidity distribution of the clusters.
Each of these cluster correlation functions satisfies 
the Thomas-Chao-Quigg relationship (with a constant $\kappa$)
$D_u(y) = \kappa (dN_{\rm ch}/dy)$ exactly.  We assume that
$\gamma_{HG} > \gamma_{QGP}$ and choose $F_{HG}$ and $F_{QGP}$ in such a way
that the QGP is concentrated near midrapidity.
A typical breakdown of the QGP and hadron gas components in our calculations
is shown in Fig.~\ref{fig:twocomp-profile}.

\begin{figure}[t]
\centerline{\includegraphics[angle=90,width=0.45\tw]{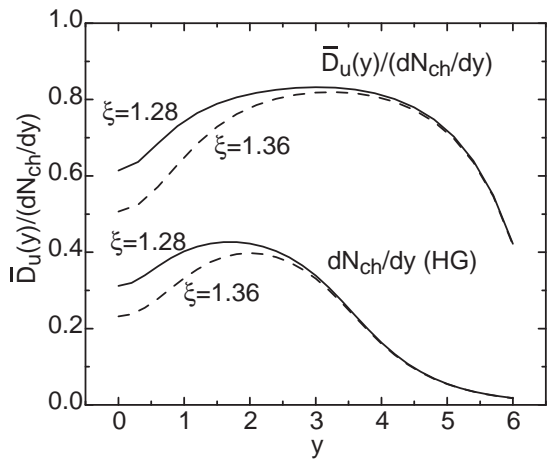}}
\caption{$D_u(y)$ as a function of $y$. $\xi$ is the size of the QGP
component. Also shown is the rapidity distribution of the charge particles
that do not belong to the QGP.}
\label{fig:figure12}
\end{figure}

If the observed rapidity window is large enough, then the observable
$\kappa(y)$ indeed shows a prominent dip  as shown in
Fig.~\ref{fig:figure12}.
Unfortunately, none of the current RHIC experiments is capable of
identifying charged particles in a large rapidity window.  Therefore we
next turn to the more realistic case of a pseudo-rapidity window of 
$-1 <\eta < 1$.  The results presented in the next section are the main
results of this study.

\section{Results for $-1 < \eta < 1$}

\begin{figure}[t]
{\includegraphics[width=0.45\tw]{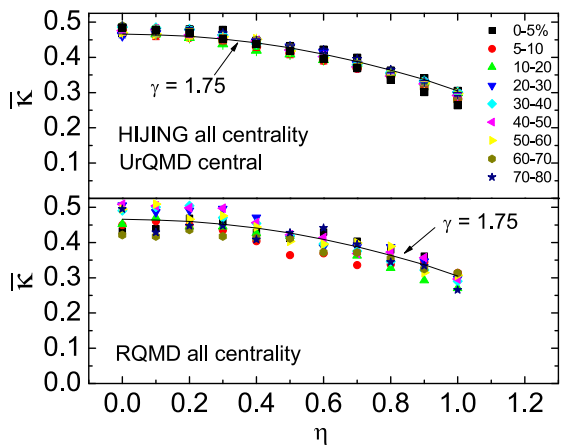}}
\caption{Results for hadronic models without a QGP within the
pseudo-rapidity range $-1 < \eta < 1$.}
\label{fig:HIJ_RQMD}
\end{figure}

To establish the baseline, we first ran 3 hadronic models without an explicit
QGP component;
 HIJING\cite{Wang:1991ht,Gyulassy:1994ew,Wang:1996yf}, 
 UrQMD\cite{Bass:1998ca,Bleicher:1999xi} and
 RQMD\cite{Sorge:1992ej,Sorge:1995dp}. 
The results are shown in
Fig.~\ref{fig:HIJ_RQMD}. As expected, $\bar{\kappa}(y)$\footnote{%
The overbars on variables indicate that these are measured only within a
finite observation window.}
does not strongly depend on the centrality in all 3 models. 
This also fixes the hadronic correlation length in our neutral cluster
model calculations to be $\gamma_{HG} = 1.75$.

\begin{figure}[t]
\centerline{\includegraphics[width=0.45\tw]{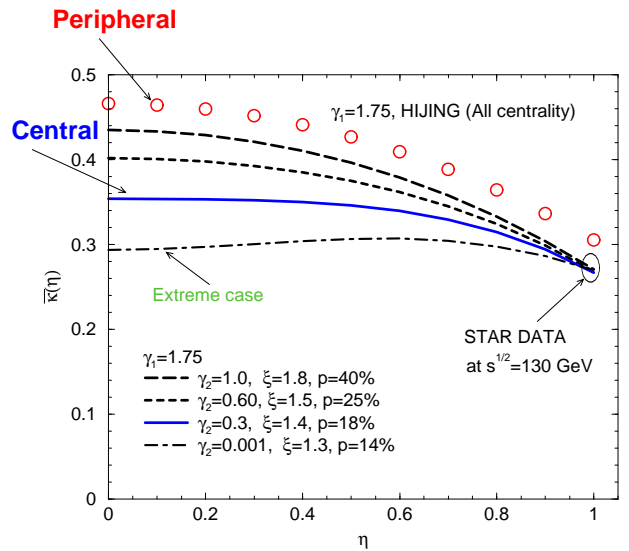}}
\caption{Results for our neutral cluster model with a QGP component
within the pseudo-rapidity range $-1 < \eta < 1$.}
\label{fig:kappa_bars2}
\end{figure}
For our neutral cluster model calculation, we first note that 
$\bar{D}_u(\eta=1)$ for the interval $-1 < \eta < 1$
is equivalent to the net charge fluctuation in the same interval. 
The experimental value for the net charge fluctuation in this rapidity
interval is known\cite{Adams:2003st}.
With the correlation length in the hadronic
part fixed at $\gamma_{HG} = 1.75$, one can vary the size of the QGP
component $\xi$ and the correlation length $\gamma_{QGP}$ within the QGP
to match $\bar{D}_u(\eta=1)$ at the known experimental value.
Then, as shown in Fig.~\ref{fig:kappa_bars2},
experimental measurements of $\bar{\kappa}(0)$ as a function of centrality
can tell us about the size of the QGP component and
the size of the charge correlation length in a QGP.
In this figure, the most reasonable scenario (in our opinion) 
for RHIC central collisions is labeled as
`Central' whereas the parametrized HIJING curve is labeled as
`Peripheral'. As one can see the values of $\bar{\kappa}(y)$
near the edge of the
observation window does not change much. This is because the amount of QGP
component at the edge is already rather small. Hence, the presence or the
absence of the QGP component makes little difference there.
On the other hand, the value of $\bar{\kappa}(y)$
at midrapidity varies as much as 30\,\%  
as the collisions become more central.

\section{Summary and Discussion}

In this article we have briefly summarized the main results of our two
papers \cite{Shi:2005rc,Jeon:2005kj}
where we have advocated the charge transfer
fluctuations as a robust signal for QGP formation.
Our observable $\kappa(y)$ is
the ratio of the charge transfer fluctuation and $dN_{\rm ch}/dy$.
We have argued that $\kappa(y)$ is in fact 
a local measure of the unlike-sign charge correlation length.  
In elementary particle collisions $(pp, K^-p)$, $\kappa(y)$ 
turned out to constant.
Hence if the charge correlation length inside QGP is
indeed small compared to the hadronic gas
\cite{Jeon:2000wg,Asakawa:2000wh,Bass:2000az}, then measuring
charge transfer fluctuations can enable us to 
detect the presence of a QGP even when the entropy fraction of the QGP is
small compared to the hadronic part.
Furthermore, by measuring how the curve $\kappa(y)$ changes from peripheral
to central collisions, one may be able to
estimate the size and the entropy fraction of the QGP component.
Current net charge fluctuations data (equivalent to
$\bar{D}_u(\eta=1)$ within $-1 < \eta < 1$) is consistent with QGP
having about $20\,\%$ of the entropy fraction and 
about 1/3 of the charge correlation length.

In view of this estimate, it is also clear why the net charge fluctuation
has not shown the expected large reduction 
\cite{Adams:2003st}:
The entropy content of QGP is small enough that if one averages over the
whole observation window, its presence is hidden behind the larger hadronic
component.

Here we would like to mention that
it will be also interesting to carry out similar studies using
strangeness and/or baryon charge.

\begin{acknowledgments}

The authors thank J.Barrette for his useful suggestions, 
V.Topor Pop for his help in running the HIJING code and also providing
the RQMD data.
The authors are supported in part by the Natural Sciences and
Engineering Research Council of Canada and by le Fonds
Nature et Technologies of Qu\'ebec.
S.J.~also
thanks RIKEN BNL Center and U.S. Department of Energy
[DE-AC02-98CH10886] for
providing facilities essential for the completion of this work.

\end{acknowledgments}


\begin{thebibliography}{17}
\expandafter\ifx\csname natexlab\endcsname\relax\def\natexlab#1{#1}\fi
\expandafter\ifx\csname bibnamefont\endcsname\relax
  \def\bibnamefont#1{#1}\fi
\expandafter\ifx\csname bibfnamefont\endcsname\relax
  \def\bibfnamefont#1{#1}\fi
\expandafter\ifx\csname citenamefont\endcsname\relax
  \def\citenamefont#1{#1}\fi
\expandafter\ifx\csname url\endcsname\relax
  \def\url#1{\texttt{#1}}\fi
\expandafter\ifx\csname urlprefix\endcsname\relax\def\urlprefix{URL }\fi
\providecommand{\bibinfo}[2]{#2}
\providecommand{\eprint}[2][]{\url{#2}}

\bibitem[{\citenamefont{Quigg and Thomas}(1973)}]{Quigg:1973wy}
\bibinfo{author}{\bibfnamefont{C.}~\bibnamefont{Quigg}} \bibnamefont{and}
  \bibinfo{author}{\bibfnamefont{G.~H.} \bibnamefont{Thomas}},
  \bibinfo{journal}{Phys. Rev.} \textbf{\bibinfo{volume}{D7}},
  \bibinfo{pages}{2752} (\bibinfo{year}{1973}).

\bibitem[{\citenamefont{Chao and Quigg}(1974)}]{Chao:1973jk}
\bibinfo{author}{\bibfnamefont{A.~W.} \bibnamefont{Chao}} \bibnamefont{and}
  \bibinfo{author}{\bibfnamefont{C.}~\bibnamefont{Quigg}},
  \bibinfo{journal}{Phys. Rev.} \textbf{\bibinfo{volume}{D9}},
  \bibinfo{pages}{2016} (\bibinfo{year}{1974}).

\bibitem[{\citenamefont{Jeon and Koch}(2000)}]{Jeon:2000wg}
\bibinfo{author}{\bibfnamefont{S.}~\bibnamefont{Jeon}} \bibnamefont{and}
  \bibinfo{author}{\bibfnamefont{V.}~\bibnamefont{Koch}},
  \bibinfo{journal}{Phys. Rev. Lett.} \textbf{\bibinfo{volume}{85}},
  \bibinfo{pages}{2076} (\bibinfo{year}{2000}), \eprint{hep-ph/0003168}.

\bibitem[{\citenamefont{Bass et~al.}(2000)\citenamefont{Bass, Danielewicz, and
  Pratt}}]{Bass:2000az}
\bibinfo{author}{\bibfnamefont{S.~A.} \bibnamefont{Bass}},
  \bibinfo{author}{\bibfnamefont{P.}~\bibnamefont{Danielewicz}},
  \bibnamefont{and} \bibinfo{author}{\bibfnamefont{S.}~\bibnamefont{Pratt}},
  \bibinfo{journal}{Phys. Rev. Lett.} \textbf{\bibinfo{volume}{85}},
  \bibinfo{pages}{2689} (\bibinfo{year}{2000}), \eprint{nucl-th/0005044}.

\bibitem[{\citenamefont{Asakawa et~al.}(2000)\citenamefont{Asakawa, Heinz, and
  Muller}}]{Asakawa:2000wh}
\bibinfo{author}{\bibfnamefont{M.}~\bibnamefont{Asakawa}},
  \bibinfo{author}{\bibfnamefont{U.~W.} \bibnamefont{Heinz}}, \bibnamefont{and}
  \bibinfo{author}{\bibfnamefont{B.}~\bibnamefont{Muller}},
  \bibinfo{journal}{Phys. Rev. Lett.} \textbf{\bibinfo{volume}{85}},
  \bibinfo{pages}{2072} (\bibinfo{year}{2000}), \eprint{hep-ph/0003169}.

\bibitem[{\citenamefont{Kafka et~al.}(1975)}]{Kafka:1975cz}
\bibinfo{author}{\bibfnamefont{T.}~\bibnamefont{Kafka}} \bibnamefont{et~al.},
  \bibinfo{journal}{Phys. Rev. Lett.} \textbf{\bibinfo{volume}{34}},
  \bibinfo{pages}{687} (\bibinfo{year}{1975}).

\bibitem[{\citenamefont{Shi and Jeon}(2005)}]{Shi:2005rc}
\bibinfo{author}{\bibfnamefont{L.-j.} \bibnamefont{Shi}} \bibnamefont{and}
  \bibinfo{author}{\bibfnamefont{S.-y.} \bibnamefont{Jeon}}
  (\bibinfo{year}{2005}), \eprint{hep-ph/0503085}.

\bibitem[{\citenamefont{Bialas et~al.}(1975)\citenamefont{Bialas, Fialkowski,
  Jezabek, and Zielinski}}]{Bialas:1974bk}
\bibinfo{author}{\bibfnamefont{A.}~\bibnamefont{Bialas}},
  \bibinfo{author}{\bibfnamefont{K.}~\bibnamefont{Fialkowski}},
  \bibinfo{author}{\bibfnamefont{M.}~\bibnamefont{Jezabek}}, \bibnamefont{and}
  \bibinfo{author}{\bibfnamefont{M.}~\bibnamefont{Zielinski}},
  \bibinfo{journal}{Acta Phys. Polon.} \textbf{\bibinfo{volume}{B6}},
  \bibinfo{pages}{39} (\bibinfo{year}{1975}).

\bibitem[{\citenamefont{Wang and Gyulassy}(1991)}]{Wang:1991ht}
\bibinfo{author}{\bibfnamefont{X.-N.} \bibnamefont{Wang}} \bibnamefont{and}
  \bibinfo{author}{\bibfnamefont{M.}~\bibnamefont{Gyulassy}},
  \bibinfo{journal}{Phys. Rev.} \textbf{\bibinfo{volume}{D44}},
  \bibinfo{pages}{3501} (\bibinfo{year}{1991}).

\bibitem[{\citenamefont{Gyulassy and Wang}(1994)}]{Gyulassy:1994ew}
\bibinfo{author}{\bibfnamefont{M.}~\bibnamefont{Gyulassy}} \bibnamefont{and}
  \bibinfo{author}{\bibfnamefont{X.-N.} \bibnamefont{Wang}},
  \bibinfo{journal}{Comput. Phys. Commun.} \textbf{\bibinfo{volume}{83}},
  \bibinfo{pages}{307} (\bibinfo{year}{1994}), \eprint{nucl-th/9502021}.

\bibitem[{\citenamefont{Wang}(1997)}]{Wang:1996yf}
\bibinfo{author}{\bibfnamefont{X.-N.} \bibnamefont{Wang}},
  \bibinfo{journal}{Phys. Rept.} \textbf{\bibinfo{volume}{280}},
  \bibinfo{pages}{287} (\bibinfo{year}{1997}), \eprint{hep-ph/9605214}.

\bibitem[{\citenamefont{Bass et~al.}(1998)}]{Bass:1998ca}
\bibinfo{author}{\bibfnamefont{S.~A.} \bibnamefont{Bass}} \bibnamefont{et~al.},
  \bibinfo{journal}{Prog. Part. Nucl. Phys.} \textbf{\bibinfo{volume}{41}},
  \bibinfo{pages}{225} (\bibinfo{year}{1998}), \eprint{nucl-th/9803035}.

\bibitem[{\citenamefont{Bleicher et~al.}(1999)}]{Bleicher:1999xi}
\bibinfo{author}{\bibfnamefont{M.}~\bibnamefont{Bleicher}}
  \bibnamefont{et~al.}, \bibinfo{journal}{J. Phys.}
  \textbf{\bibinfo{volume}{G25}}, \bibinfo{pages}{1859} (\bibinfo{year}{1999}),
  \eprint{hep-ph/9909407}.

\bibitem[{\citenamefont{Sorge et~al.}(1992)\citenamefont{Sorge, Berenguer,
  Stocker, and Greiner}}]{Sorge:1992ej}
\bibinfo{author}{\bibfnamefont{H.}~\bibnamefont{Sorge}},
  \bibinfo{author}{\bibfnamefont{M.}~\bibnamefont{Berenguer}},
  \bibinfo{author}{\bibfnamefont{H.}~\bibnamefont{Stocker}}, \bibnamefont{and}
  \bibinfo{author}{\bibfnamefont{W.}~\bibnamefont{Greiner}},
  \bibinfo{journal}{Phys. Lett.} \textbf{\bibinfo{volume}{B289}},
  \bibinfo{pages}{6} (\bibinfo{year}{1992}).

\bibitem[{\citenamefont{Sorge}(1995)}]{Sorge:1995dp}
\bibinfo{author}{\bibfnamefont{H.}~\bibnamefont{Sorge}},
  \bibinfo{journal}{Phys. Rev.} \textbf{\bibinfo{volume}{C52}},
  \bibinfo{pages}{3291} (\bibinfo{year}{1995}), \eprint{nucl-th/9509007}.

\bibitem[{\citenamefont{Adams et~al.}(2003)}]{Adams:2003st}
\bibinfo{author}{\bibfnamefont{J.}~\bibnamefont{Adams}} \bibnamefont{et~al.}
  (\bibinfo{collaboration}{STAR}), \bibinfo{journal}{Phys. Rev.}
  \textbf{\bibinfo{volume}{C68}}, \bibinfo{pages}{044905}
  (\bibinfo{year}{2003}), \eprint{nucl-ex/0307007}.

\bibitem[{\citenamefont{Jeon et~al.}(2005)\citenamefont{Jeon, Shi, and
  Bleicher}}]{Jeon:2005kj}
\bibinfo{author}{\bibfnamefont{S.}~\bibnamefont{Jeon}},
  \bibinfo{author}{\bibfnamefont{L.}~\bibnamefont{Shi}}, \bibnamefont{and}
  \bibinfo{author}{\bibfnamefont{M.}~\bibnamefont{Bleicher}}
  (\bibinfo{year}{2005}), \eprint{nucl-th/0506025}.

\end{thebibliography}
\end{document}